# Neutron-Irradiation Induced Magnetization and Persistent Defects at High Temperatures in Graphite

Ranjan Mittal[1,2], Mayanak K. Gupta[1], Sanjay K. Mishra[1], Sourabh Wajhal[1], Peram Delli Babu[3], Baltej Singh[1,2], Anil Bhimrao Shinde[1], Poluri Siva Rama Krishna[1], Rakesh K. Singhal[4], Rakesh Ranjan[5] and Samrath Lal Chaplot[1,2]

[1]*Solid State Physics Division, Bhabha Atomic Research Centre, Mumbai, 400085, India*
[2]*Homi Bhabha National Institute, Anushaktinagar, Mumbai, 400094, India*
[3]*UGC-DAE Consortium for Scientific Research, Mumbai Centre, R5-Shed, BARC, Trombay, Mumbai, 400 085, India*
[4]*Analytical Chemistry Division, Bhabha Atomic Research Centre, Mumbai, 400085, India*
[5]*Reactor Operations Division, Bhabha Atomic Research Centre, Mumbai, 400085, India*

Structural as well as magnetization studies have been carried out on graphite samples irradiated by neutrons over 50 years in the CIRUS research reactor at Trombay. Neutron diffraction studies reveal that the defects in irradiated graphite samples are not well annealed and remain significant up to high temperatures much greater than 653 K where the Wigner energy is completely released. We infer that the remnant defects may be intralayer Frenkel defects, which do not store large energy, unlike the interlayer Frenkel defects that store the Wigner energy. Magnetization studies on the irradiated graphite show ferromagnetic behavior even at 300 K and a large additional paramagnetic contribution at 5 K. Ab-initio calculations based on the spin-polarized density-functional theory show that the magnetism in defected graphite is essentially confined on to a single 2-coordinated carbon atom that is located around a vacancy in the hexagonal layer.

Graphite is used as a moderator and reflector material in nuclear fission reactors. Graphite [1] also serves as a non-replaceable structural component in reactor cores, which must remain stable over the lifetime of the reactor while operating at elevated temperatures. In a reactor, neutron irradiation damage in graphite results in the formation of structural defects, such as displaced atoms, vacancies, and higher order aggregates that raise the internal energy. Relaxation of the defects leads to spontaneous release of energy at around 473 K, which is referred as Wigner energy [2]. This stored energy associated with defects can surprisingly reach large energy densities with values as high as 2.7 MJ/kg. The bursts of released energies may be intense enough to cause fires in reactor and lead to nuclear accident. Hence fundamental studies of the defect physics in graphite is important to assess safety aspects concerning to reactor decommissioning. The understanding of the structure and dynamics at the atomic level in graphite, as well as in other graphitic materials such as graphene and carbon nanotubes, subjected to the intense radiation and high temperatures is of fundamental interest [3-17].

Ab-initio calculations of the structure, energies and behaviour of defects in graphitic materials have been reported [4,17-19]. We have recently reported our experimental and ab-initio computational studies on characterization of neutron-irradiation induced defects in graphite and the mechanism of release of the Wigner energy [20]. The magnetic behavior has been previously investigated experimentally on unirradiated samples of highly-oriented pyrolytic graphite (HOPG) and also those irradiated with low neutron fluence ($3.12 \times 10^{19}$ neutrons/cm$^2$) [21-23], protons and Helium atoms [24,25].

Here we focus our studies to understand the effect of irradiation on electronic, magnetic, and structural properties of graphite. We used samples [26] irradiated at various levels of neutron fluence in the reflector section of the CIRUS research reactor at Trombay, India. The highest fluence of the neutrons encountered by the samples is $2.6 \times 10^{21}$ neutrons/cm$^2$. The values of displacement per atom (dpa) for neutron energy more than 50 KeV has been obtained by multiplying with a factor of $0.68 \times 10^{-21}$ dpa.cm$^2$/neutrons to the neutrons fluence, following Ref. [27]. The details of various samples are given in **TABLE SI** in Supplementary Material [28].

We report neutron diffraction studies on these irradiated samples as a function of temperature, and find that the structure continues to be significantly defected even at very high temperatures much above the temperature where the Wigner energy is fully released. We have first time studied magnetization behavior of highly irradiated graphite samples. Our samples involving significant atomic-level defects are different from HOPG samples, which essentially involve orientational disorder of graphitic layers. Our measurements indicate that a low neutron fluence does not alter the magnetic properties of graphite. i.e. it remains in the nonmagnetic (or diamagnetic phase), while above certain irradiation level graphite starts exhibiting a weak magnetic moment. It should be noted that HOPG samples showed magnetization behavior with both the unirradiated and irradiated samples.

**Persistent Defects at High Temperatures**

The temperature of the inner region of graphite reflector of CIRUS reactor was in between 323 to 403 K over the operation of the reactor over five decades. The average temperature as seen by graphite is about 363 K. Differential scanning calorimeter measurements (**Fig. S1**, Supplementary Material [28]) for these samples show that the onset temperature for release of Wigner energy is 408 K, while Wigner energy is completely released at 653 K.

The microscopic changes in the structure can be understood from diffraction studies. In order to see the effect of annealing on the structure of irradiated samples, we performed temperature-dependent neutron diffraction measurements from 300 K to 738 K. **Fig. 1** shows typical neutron diffraction data of the highly irradiated and unirradiated graphite samples at 738 K. To quantify the changes in the lattice parameters with temperature, we performed the Rietveld refinement of the powder neutron diffraction patterns (**Fig. 1**) using the hexagonal structure with space group: P6$_3$/mmc. Evolution of the powder neutron diffraction pattern of unirradiated and irradiated graphite with temperature and Rietveld refinement of the powder neutron diffraction patterns at several temperatures are given in **Fig. S2** and **S3** in Supplementary Material [28].

Structural parameters obtained from the Rietveld refinement from 300 K to 738 K of both the samples are shown in **Fig 2**. Graphite is known to show highly anisotropic thermal-expansion behavior [29-35]. For the unirradiated sample, we find that on increasing the temperature, the $c$ lattice parameter shows large expansion and $a$ lattice parameter shows a very small contraction. This is consistent with the nature of bonding in graphite. Graphite is known to have van der Waals bonding along the $c$-axis, which is much weaker than the covalent bond in the $a$-$b$ plane. The Rietveld analysis of the diffraction patterns shows (**Fig 1**) that on increasing the temperature both the samples show that lattice expands along the $c$-axis but with different magnitudes.

The expansion in the $c$-direction on increase of temperature from 300 K to 738 K for the unirradiated and irradiated samples is 0.1 Å and 0.05 Å, respectively. The expansion in the irradiated sample is found to be significantly less compared to that in the unirradiated sample. The expansion along the $a$-axis is observed to be negative in the unirradiated sample. In contrast to this, the lattice expansion along $a$- axis in the irradiated sample is positive i.e. it expands on heating.

The above neutron diffraction results clearly show that even after heating the sample at 738 K (significantly well above the complete Wigner energy release point of 653 K), the disorder in the system persists. We can try to understand the nature of remnant defects as follows. We note that the expansion in the c-lattice parameter, which reflects the interlayer spacing, is much smaller in the irradiated sample than in the unirradiated sample. This indicates that the interstitial atoms have returned to the layer, which helps to lower the expansion. However, the in-plane lattice parameter has expanded much more in the irradiated sample. This suggests that the returned interstitial atom has not yet combined with the vacancy; that is, the Frenkel defect has not fully annealed. This interpretation of a metastable Frenkel defect within a layer is consistent with our previous ab-initio molecular dynamics simulations [20], where we found that, prior to the full annealing at 1000 K, the simulation at 900 K showed a metastable defect structure consisting of five- and seven-member carbon rings within a graphite layer (**Fig. 2**). The structure did not relax further into hexagonal rings in fairly long simulation time. The simulations [20] show that such an intralayer defect stores about 5 eV of energy, while the interlayer defect stores around 15 eV. It is likely that the concentration of the intralayer defects is not vary large, and so the net energy in the remnant defects at high temperatures is negligible compared to the large Wigner energy already released.

It can be seen that the lattice parameters in the temperature range from 420 K to 550 K show large variations (**Fig. 2**). This may be related to the temperature range for the release of the Wigner energy in our irradiated sample of graphite[20]. It can be seen that in this temperature range we have collected the diffraction data at large number of temperatures. As the Wigner energy is gradually released on increase of the temperature, there may be local temperature fluctuations within the sample, which might cause the variations of the average lattice parameters.

The large anisotropy in the graphite crystal structure is also observed in the anisotropic thermal parameters' values as obtained from the Rietveld refinements of the neutron diffraction data of both the samples shown in **Fig. 2**. The thermal parameters along the $c$-axis are larger than those in the $a$-$b$ plane, which is expected from the weaker van-der-Waals bonding along the $c$-axis than the covalent bonding within the $a$-$b$ plane. So also, the thermal parameters for the irradiated sample (**Fig. 2**) are larger than those for the un-irradiated sample, which reflect the static positional disorder due to the defects. The excess values in $U_{11}$ and $U_{33}$ thermal parameters are about 0.003 and 0.01 Å$^2$, which corresponds to root-mean-amplitude of the disorder of about 0.05 and 0.1 Å in the $a$-$b$ plane and along the $c$-axis respectively. These values are in fairly good agreement with the positional disorder found in our ab-initio simulation of the defected structure with a Frenkel defect.

**Neutron-Irradiation Induced Magnetization**

We performed magnetization measurements on four different irradiated and unirradiated graphite samples (TABLE SI [28]). The samples S0 (0.0645 dpa) and S4 (0.0125 dpa) are highly irradiated, while S11 (0.00004 dpa) and S12 (unirradiated) have seen minimal or no irradiation. The raw (uncorrected) magnetization data are given in **Fig. S4** (Supplementary Material [28]). The S11 and S12 samples exhibit essentially pure diamagnetic character in the M vs H curves both at 5 K and 300 K. The magnetization data for the S0 and S4 samples, corrected for the diamagnetic contribution coming from the sample, are shown in **Fig. 3**. The zoomed portion near the origin shown in the lower inset of **Fig. 3** clearly shows presence of ferromagnetic (FM) like magnetic order even at 300 K for these two samples. The upper inset shows further zooming around the origin, which reveals the loop opening more clearly. The presence of magnetic hysteresis and coercivity ($H_C$) are clearly seen for both these samples with $H_C$ values of ~ 250 Oe at 5K and ~110 Oe at 300 K for both S0 and S4. The saturation magnetization values are 0.28 emu/g (0.11 emu/g) at 5 K and $4.4\times10^{-3}$ emu/g ($4.8\times10^{-3}$ emu/g) at 300 K for S0 (S4) sample. On the other hand, the unirradiated S12 or the least irradiated S11 samples, do not show any signature of magnetism. This clearly indicates that the weak magnetism observed for the highly irradiated S0 and S4 samples is essentially arising from irradiation effects.

Another point to note is that there seems to be a large paramagnetic contribution at 5 K in both S0 and S4 samples in addition to FM like contribution noticed at 300 K. Irradiation seems to have broken many chemical bonds resulting in significant number of unpaired spins. For a long-range magnetic order to occur, in addition to the unpaired spins i.e. magnetic moments, an interaction between them is necessary. In the present samples, a weak interaction between spins seems to order them ferromagnetically, and many other spins remain in paramagnetic (PM) state. As the temperature is lowered some correlations between these spins increase, and hence, the magnetization increases at 5 K. The spin system in the sample on the whole exhibits FM+PM like behavior which is different than the previously reported ferromagnetic [21,23] and paramagnetic [22] behavior for unirradiated and low neutron irradiated HOPG samples respectively. On the other hand, hydrogen and helium ions irradiation on HOPG samples [24] showed ferromagnetic behavior, while proton irradiation revealed [25] ferro- or ferrimagnetism.

A weak FM behavior was observed [21] in magnetization studies on HOPG, Krish graphite and natural graphite samples. The origin of FM behavior in Krish and natural graphite samples was attributed to iron impurities. However, the impurity contribution was ruled out in case of HOPG. In the present case, the magnetism due to the presence of any magnetic impurities can be ruled out as the unirradiated and the least irradiated samples show only diamagnetic behavior. Further, the S0 sample, which has received highest radiation dose, exhibits stronger magnetic behavior than that in the sample with a lower dose (S4). All these facts put together lead to the conclusion that the magnetism observed in the irradiated samples arises mainly due to defects induced by irradiation.

The analysis of Fe impurities in our graphite samples is performed by complete dissolution of the material [36]. The concentration of Fe in our samples is found to be as low as 8.0 ± 0.4 ppm. An impurity concentration of 1 ppm in graphite is expected [21] to contribute $2.2\times10^{-4}$ emu/g to the magnetization in case of the impurity as Fe clusters, In case the impurity is in the form of $Fe_3O_4$ clusters, the contribution from 1 ppm impurity to magnetization would be $1.4\times10^{-4}$ emu/g. If we assume that a linear increase of the magnetic moment with the Fe concentration, the saturation magnetization values of 0.28 emu/g, (0.11 emu/g) at 5 K and $4.4\times10^{-3}$ emu/g ($4.8\times10^{-3}$ emu/g) at 300 K for S0 (S4) sample are much more than the likely contribution of $\sim10^{-3}$ emu/g from 8 ppm of Fe impurity in our samples.

In order to understand the origin of magnetization in graphite, we have performed ab-initio spin-polarized density-functional theory calculations. Details of the method are given in the Supplementary Material [28]. The minimum energy structure of the perfect crystal does not show any magnetism. We then simulated the graphite structure with a Frenkel defect. The Frenkel defect was created by removing one carbon atom from one layer and placed at a position away from the vacant site between the two graphite layers in the 4×4×1 supercell containing 64 atoms. After relaxation, the minimum energy structure (**Fig 4(a)**) with a Frenkel defect is found to have 2-coordinated carbon atoms around the vacancy and 4-coordinated carbon atoms bonded to the interstitial atom. A vacancy in the layer leads to formation of three distorted pentagons with each having 2-coordinated atoms (**Fig 4(a)**). We find that one of these 2-coordinated carbon atoms is magnetic with a moment value of 0.51 $\mu_B$. The calculated magnetic moment on various atoms is shown in **Fig.4(a).** The total magnetic moment associated with the Frenkel defect is found to be 0.76 $\mu_B$. With the one Frenkel defect in the 4×4×1 supercell, the magnetic moment per unit volume, i.e., the spin density = 0.76/V = $1.3\times10^{-3}$ $\mu_B$·Å$^{-3}$; (V is the supercell volume). The value of the magnetic moment due to a Frenkel defect in graphite in our calculation is comparable to the value of 1.04 $\mu_B$ found around a vacancy in a graphene sheet in a previous ab-initio calculation [4].

The origin of magnetism in our present studies on irradiated graphite and previous studies on HOPG samples [21-23] seems to be different. We note that the perfect graphite structure, as well as a single graphene layer, do not show any magnetism. In irradiated graphite the magnetism is found to be around a 2-coordinated carbon atom that is located close to the vacancy. In case of HOPG, the magnetism seems to arise from orientational disorder of the layers and interlayer interactions.

Further, we have performed the electronic density of states calculation using the spin-polarized DFT method. The finite density of states at the Fermi level (**Fig 4(b)**) in pure graphite is contributed from one free electron after sp$^2$ hybridization. This free electron is delocalized within a graphite layer, and does not contribute to any net magnetic moment. However, introduction of defect in graphite leads to magnetism as discussed above. Our DFT calculation in defected graphite shows significant value of the density of states at the Fermi energy (**Fig 4(b)**), which suggest that some new electronic states have been created due to the Frenkel defect formation that participate in electronic conduction. The new conducting states are mainly formed due to the 2-coordinated carbon atoms in the vicinity of the vacancy defect (**Fig 4(b)**), which is also reflected in the additional charge density around the vacancy in the charge density plot (**Fig. 4(c)**).

We have reported the results of neutron diffraction and magnetization experiments on highly irradiated graphite samples from the reflector region of CIRUS research reactor at Trombay. We may note that other graphitic materials such as graphene and carbon nanotubes have a two-dimensional layered structure similar to that of graphite. Thus, our studies performed on irradiated graphite samples are also applicable to other graphitic materials. The structural studies on irradiated graphite samples show that the static disorder along the *c*-axis of the graphite hexagonal structure persists at high temperatures much above the temperature of release of the Wigner energy (653 K). The persistence of the defects at high temperatures has important consequences for the application of graphitic materials in high radiation environments. The magnetization experiments reveal weak ferromagnetism at 300 K, and also a large paramagnetic contribution at low temperature of 5 K. Ab-initio spin-polarized density-functional calculations are performed to understand the origin of magnetism in irradiated graphite, which show that the magnetism arises on one of the 2-coordinated carbon atoms in the region around a vacancy. Our experimental observation of magnetism in defected graphitic materials is of fundamental importance as these materials find wide applications.

## ACKNOWLEDGEMENTS
The use of ANUPAM super-computing facility at BARC is acknowledged. SLC acknowledges the financial support of the Indian National Science Academy for INSA Senior Scientist position.

FIG 1 (Color online) The Rietveld refinement of neutron diffraction patterns for highly irradiated graphite sample S0 (0.0645 dpa) and another unirradiated graphite sample S12 at 738 K.

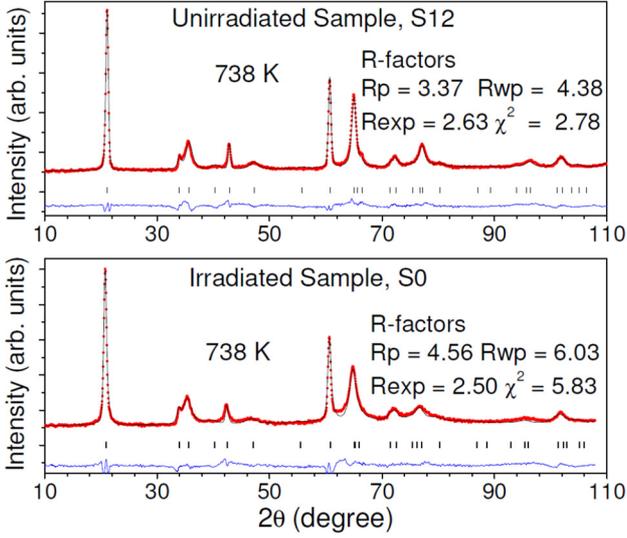

FIG 2 (Color online) Variation of the structural parameters (lattice and thermal parameters) of irradiated and unirradiated graphite as obtained from Rietveld analyses of neutron data. The graphite sample S0 (0.0645 dpa) was irradiated with neutrons, and S12 was unirradiated. We show a structure with an intralayer Frenkel defect consisting of five- and seven-member carbon rings as found in AIMD simulation at 900 K [20].

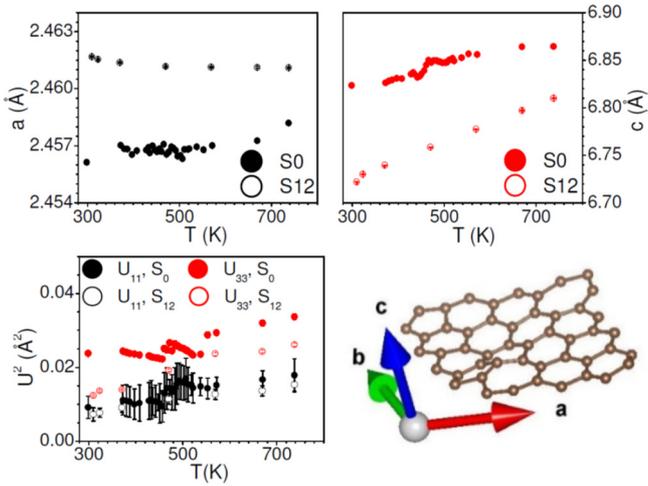

FIG 3 (Color online) Magnetization versus field at 5K and 300 K for highly irradiated graphite samples, S0 (0.0645 dpa) and S4 (0.0125 dpa). These data have been corrected for diamagnetic contribution as described in the text. Both the insets show MH curves close to the origin with different amounts of zooms. The lower inset brings out the ferromagnetic-like magnetic order present even at 300K. The upper inset shows the presence of coercivity at 5 K and 300 K.

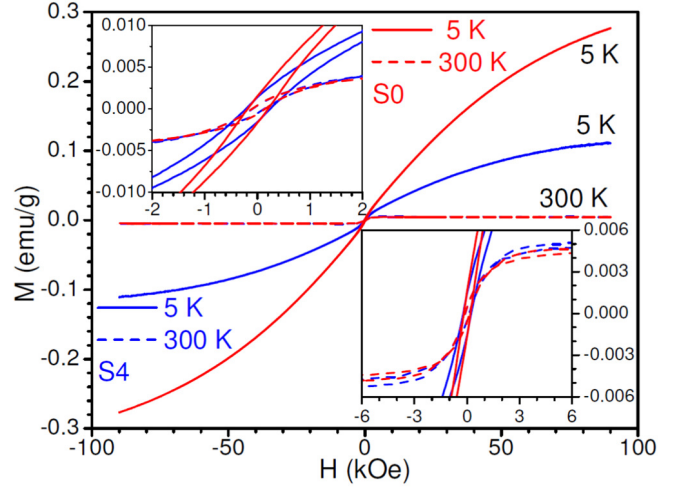

FIG. 4 (Color online) (a) A graphite layer with a single Frenkel defect in a 4×4×1 supercell. Only one atom (red color) shows significant value of magnetic moment 0.51 $\mu_B$; a few atoms (green color) have very small moment of 0.03 ±0.02 $\mu_B$ and the rest of the atoms (blue color) have nearly 0 ± 0.01 $\mu_B$. (b) The calculated electronic density of states of graphite with a single Frenkel defect in a 4×4×1 supercell and with no defect. (c) The calculated charge density (e/Å$^3$) of graphite (Left) with a single Frenkel defect in a 4×4×1 supercell and (Right) with no defect.

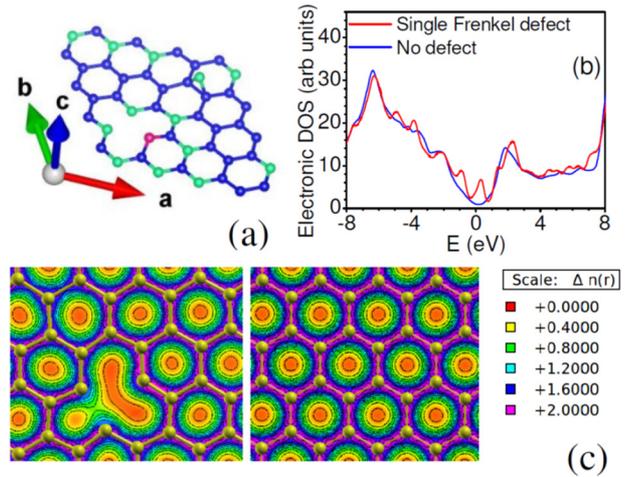

# Supplementary Materials

# Neutron-Irradiation Induced Magnetization and Persistent Defects at High Temperatures in Graphite

Ranjan Mittal[1,2], Mayanak K. Gupta[1], Sanjay K. Mishra[1], Sourabh Wajhal[1], Peram Delli Babu[3], Baltej Singh[1,2], Anil Bhimrao Shinde[1], Poluri Siva Rama Krishna[1], Rakesh K. Singhal[4], Rakesh Ranjan[5] and Samrath Lal Chaplot[1,2]

[1]*Solid State Physics Division, Bhabha Atomic Research Centre, Mumbai, 400085, India*
[2]*Homi Bhabha National Institute, Anushaktinagar, Mumbai 400094, India*
[3]*UGC-DAE Consortium for Scientific Research, Mumbai Centre, R5-Shed, BARC, Trombay, Mumbai - 400 085, India*
[4]*Analytical Chemistry Division, Bhabha Atomic Research Centre, Mumbai, 400085, India*
[5]*Reactor Operations Division, Bhabha Atomic Research Centre, Mumbai, 400085, India*

## EXPERIMENTAL

The graphite samples [1] irradiated with neutron fluence at various levels were taken out from a block of irradiated graphite originally used in the reflector section of the CIRUS research reactor at Trombay. The samples are designated as S0, S1, --- S11, which have been irradiated with decreasing level of neutron flux as given Table 1. We have also used an irradiated sample for reference. These samples are characterized [2] by X-ray and neutron diffraction techniques, differential scanning calorimetry and small-angle X-ray scattering. The neutron powder diffraction data have been recorded in the $2\theta$ range of $4°$–$138°$ in steps of $0.1°$ using neutrons of wavelength of 1.2443 Å on the powder diffractometer PD2 at the Dhruva reactor at Bhabha Atomic Research Centre at Trombay, Mumbai. An 8 mm diameter thin-walled can made of vanadium niobium alloy was used to hold the polycrystalline sample of graphite. The high temperature neutron diffraction studies from 300 to 738 K are performed using a furnace. The vanadium normalization and background subtraction has been performed using the in-house developed codes specific to the instrument. The analysis of neutron diffraction data are performed using the Rietveld [3] refinement method using FULLPROF [4] program.

The magnetization measurements were carried out using a commercial (Quantum Design Make) 9 T Physical Property Measurement system (PPMS) based Vibrating Sample Magnetometer (VSM).

The analysis of Fe impurities in our graphite samples is carried out by complete dissolution of the material [5]. Before dissolution, a sample of 0.5 g is dry washed at 800°C for 2 hr. Then the residue is dissolved in 2% nitric acid. The analysis of Fe in completely dissolved sample is done by using the inductively-coupled plasma optical emission spectrometry (ICP-OES, make Thermofisher model I CAP 6000 series). Concentration of Fe is measured by using an emission line at 259.9 nm. The calibration of the equipment was done by using a standard solution of Fe having a strength of 1000 µg mL$^{-1}$. The concentration of Fe is found to be as low as $8.0 \pm 0.4$ ppm.

## COMPUTATIONAL DETAILS

The ab-initio calculations are performed Vienna-based ab-initio simulation package (VASP) [6,7]. The force calculation of distinct configuration are performed using the Vienna based ab-initio simulation package[6,7] (VASP). The exchange correlation part of DFT formalism is treated with generalized gradient approximation with Perdew, Burke and Ernzerhof [8,9] (PBE) functional. Further to incorporate the interaction between the valance and core electrons the projected augmented wave method is used. A plane wave basis set cutoff of 1000 eV was used for basis set expansion. The Brillouin zone integrations were performed on 20×20×5 mesh generated using the Monkhorst-Pack method[10]. The electronic convergence and ionic forces tolerance were chosen to be $10^{-8}$ eV and $10^{-3}$ eV/Å respectively. All the calculations are performed with relaxed geometry. The relaxation of the graphite structure was performed with including various vdW-DFT non-local correlation functional [11-13]. We find that the relaxation performed using the optB88-vdW [13] (BO) functional scheme of the vdW-DFT method produces the best match with the experimental structure.

The structure relaxation in a 4×4×1 supercell containing 64 atoms with a single Frenkel defect is performed in various magnetic configurations to estimate the magnetic moment in the defect structure. The Frenkel defect was created by removing one carbon atom from one layer and placed at a position away from the vacant site between the two graphite layers in the 4×4×1 supercell. Several computational possible schemes of introducing antiferromagnetic (A-Type, C-Type and G-Type) and ferromagnetic ordering were considered while doing the structure relaxation. However, due to the large separation between the layers along the c-axis, the structure relaxation with various magnetic schemes mentioned above always resulted in a similar magnetic structure.

TABLE SI. Neutron fluence as seen by various graphite samples [1]. Another unirradiated sample for reference is assigned as sample number S12. The fluence of neutrons of energy more than 50 KeV is converted to displacement per atom (dpa) by multiplying [14] by a factor of 0.68 x 10$^{-21}$ dpa.cm$^2$/neutrons. The displacement per atom (dpa) is given in the units of 10$^{-3}$. The samples are located at various radial distances away from the reactor core. The radial distances of various graphite samples are with respect to the sample S0 which had seen the largest neutron fluence. The TABLE given here is taken from our previous publication [2].

| Sample No. | Radial distance (cm) | Thermal neutrons (×10$^{20}$ neutrons/cm$^2$) (E < 0.625 eV) | Epithermal neutrons, (×10$^{19}$ neutrons/cm$^2$) (0.625 eV < E < 0.82 MeV) | Fast neutrons, (×10$^{18}$ neutrons/cm$^2$) (E > 0.82 MeV) | Displacement per atom (dpa) (×10$^{-3}$) |
|---|---|---|---|---|---|
| S0 | 0 | 22.06 | 37.69 | 19.99 | 64.5 |
| S4 | 20 | 15.09 | 7.38 | 3.70 | 12.5 |
| S11 | 90 | 0.75 | 0.03 | 0.0013 | 0.04 |

FIG. S1 (Color online) Heat flow versus temperature for graphite sample irradiated with different fluence of neutron [2]. The neutron fluence seen by the samples is given in TABLE SI. Another un-irradiated sample for reference is assigned as sample number S12. This figure has been taken from our previous publication [2].

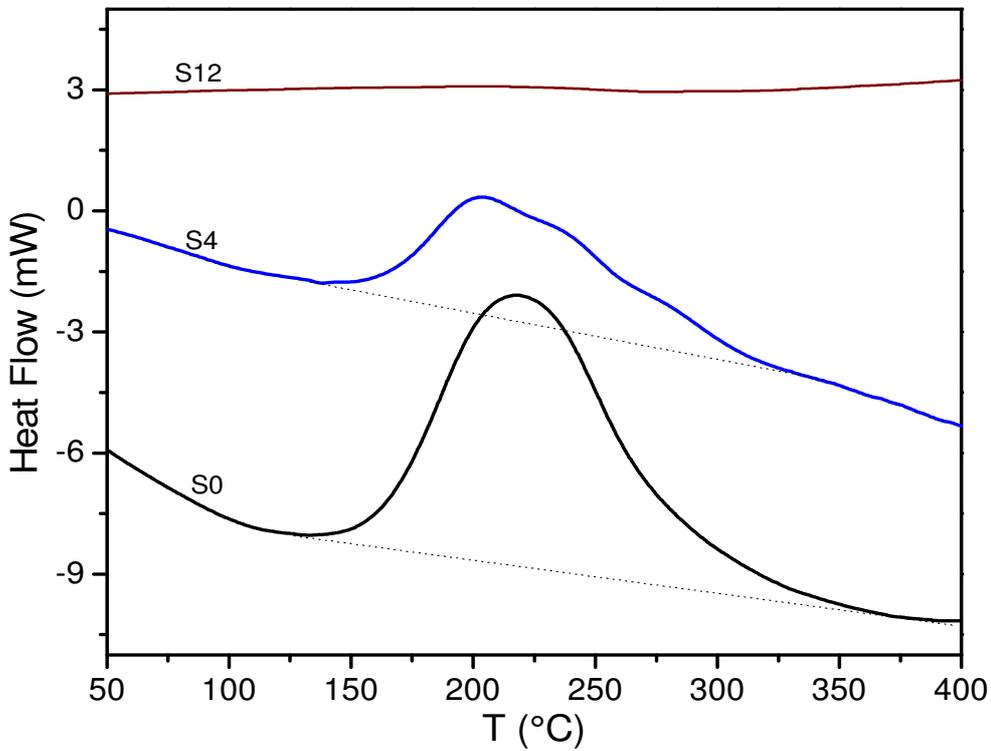

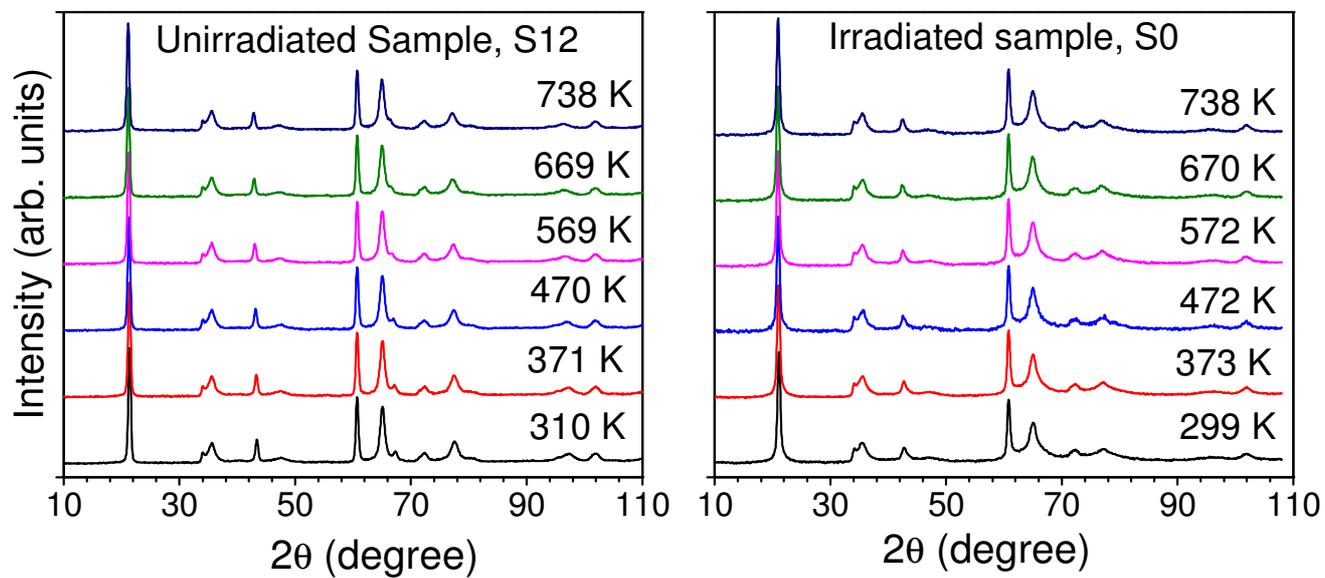

FIG S2 (Color online) Evolution of the powder neutron diffraction pattern of highly irradiated graphite sample S0 (0.0645 dpa) and another unirradiated graphite sample S12 with temperature.

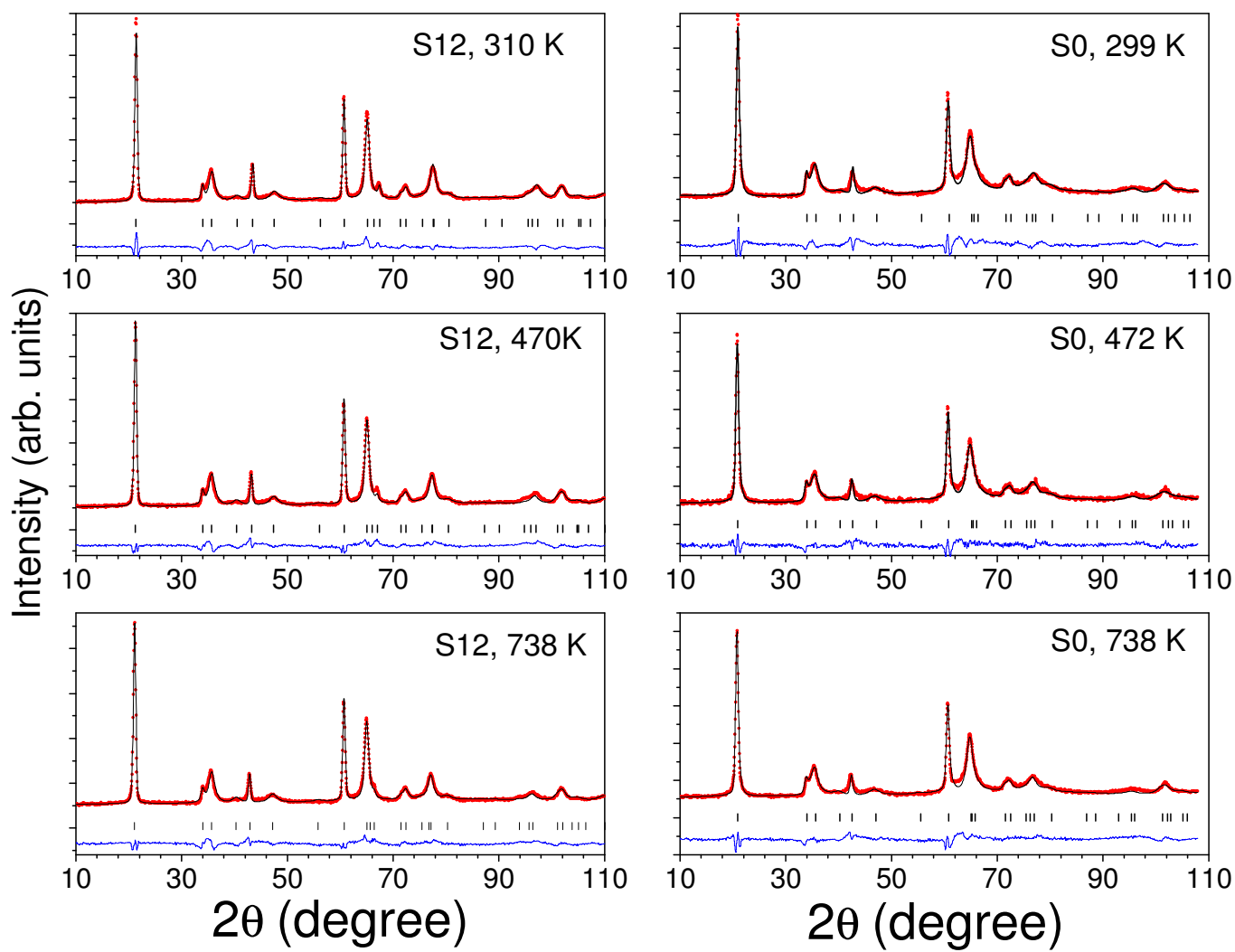

FIG S3 (Color online) The Rietveld refined neutron diffraction patterns for highly irradiated graphite sample S0 (0.0645 dpa) and another unirradiated graphite sample S12.

FIG S4 (Color online) As measured magnetization versus field for S0, S4, S11 and S12 samples at 5 K and 300 K. The samples S0 (0.0645 dpa) and S4 (0.0125 dpa) are highly irradiated, while S11 (0.00004 dpa) and S12 (unirradiated) have seen minimal or no irradiation. For clarity, refer TABLE SI.

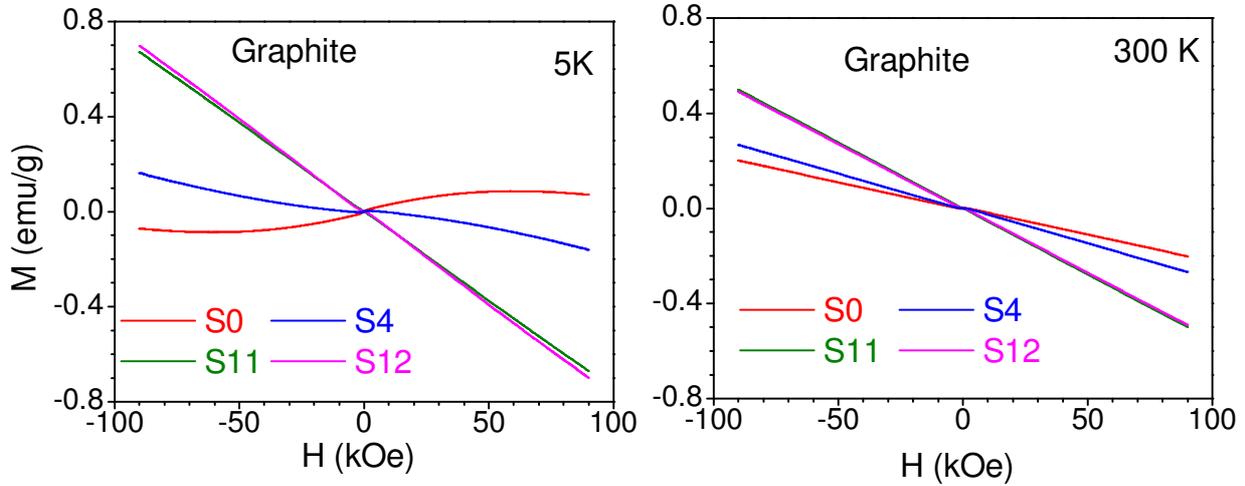